\documentclass[conference]{IEEEtran}
\IEEEoverridecommandlockouts

\usepackage[]{inputenc}
\usepackage{esvect}
\usepackage[sort]{cite}
\usepackage{longtable}
\usepackage[colorlinks=true,citecolor=blue]{hyperref}
\usepackage{fouriernc}
\usepackage{physics}
\usepackage{graphicx}
\usepackage{bm}
\usepackage{qcircuit}
\usepackage{ragged2e}
\usepackage{todonotes}
\usepackage{comment}
\usepackage{amstext}
\usepackage{subcaption}
\usepackage{amsmath,amssymb,amsfonts}
\usepackage{algorithm}
\usepackage{framed}
\usepackage{bbm}
\usepackage{algpseudocode}
\usepackage{textcomp}
\usepackage{xcolor}
\usepackage{orcidlink}
\def\BibTeX{{\rm B\kern-.05em{\sc i\kern-.025em b}\kern-.08em
    T\kern-.1667em\lower.7ex\hbox{E}\kern-.125emX}}
\begin{document}


\title{Efficient variational quantum eigensolver methodologies on quantum processors}

\author{\IEEEauthorblockN{Tushar Pandey}
\IEEEauthorblockA{
\textit{Texas A\&M University}\\
College Station, Texas \\
tusharp@tamu.edu}\orcidlink{0000-0001-7448-5723}
\and
\IEEEauthorblockN{Jason Saroni}
\IEEEauthorblockA{\textit{Iowa State University} \\
\textit{Ames, Iowa}\\
jsaroni@iastate.edu}\orcidlink{0009-0003-7197-2309}
\and
\IEEEauthorblockN{Abdullah Kazi}
\IEEEauthorblockA{
\textit{University of Mumbai
}\\
India \\
Abdullah.k@gmail.com}
\and
\IEEEauthorblockN{Kartik Sharma}
\IEEEauthorblockA{
\textit{Indian Institute of Science}\\
India \\
kartiksharma@iisc.ac.in}
}

\maketitle

\begin{abstract} 
 \noindent We compare the performance of different methodologies for finding the ground state of the molecule $BeH_2$. We implement adaptive, tetris-adaptive variational quantum eigensolver (VQE), and entanglement forging to reduce computational resource requirements. We run VQE experiments on IBM quantum processing units and use error mitigation, including twirled readout error extinction (TREX) and zero-noise extrapolation (ZNE) to reduce noise. Our results affirm the usefulness of VQE on noisy quantum hardware and pave the way for the usage of VQE related methods for large molecules.
 \end{abstract}


\section{Introduction}
VQE is a hybrid algorithm that combines quantum and classical computing to find the ground state of a system, a key task in quantum chemistry. It uses a parameterized quantum circuit to prepare a quantum state, and a classical optimizer to adjust the circuit parameters. The expectation value of the system's Hamiltonian represents the total energy. This method has been successfully applied to find the ground state energy of various molecules, leveraging the strengths of both quantum and classical computing to tackle this complex problem. VQE's are especially promising for implementation on Noisy Intermediate-Scale Quantum (NISQ) devices and can be tailored to reduce circuit depth and used qubits.

The VQE's application to $BeH_2$ is challenging due to the molecule’s complexity and the size of quantum system required as the molecule is relatively large compared to extensively studied molecules like $H_2$ and $LiH$ \cite{10.1038/nature23879}. The transformation of the molecule’s Hamiltonian to a qubit Hamiltonian, which involves simulating the electronic orbital interactions, is complex. Additionally, the increasing circuit depth for larger molecules like $BeH_2$ presents a significant challenge. However knowing the ground state energy of a molecule has a range of applications including helping to predict its reactivity, chemical stability, and spectroscopic properties. With this motivation in mind, we perform resource efficient executions of VQE. We use adaptive and tetris-adaptive VQE to reduce the number of two qubit gates and circuit depth. Sec.~\ref{sec:methodologies} describes the mapping of the VQE problem onto a quantum computer and the error mitigation strategies and techniques we implement. Sec.~\ref{sec:results} presents the results from adaptive and tetris-adaptive VQE and from running our VQE algorithm on IBM hardware with TREX and ZNE. Sec.~\ref{sec:discussion} discusses the results in light of the theoretical expectation, and finally Sec.~\ref{sec:conclusion} the conclusion and future prospects.

\section{Methodologies}
\label{sec:methodologies}
\subsection{Variational Quantum Eigensolver}
VQE is a quantum algorithm designed to estimate the ground state energy of a quantum system using quantum computers.
\begin{equation}
    H |\Phi \rangle = E_G |\Phi \rangle,
\end{equation}
where $H$ obtained using the Born-Oppenheimer approximation, is the non-relativistic time-independed hamiltonian that models the constituents of the molecules. In first quantized form,

\begin{align}
H = -\sum_{i}\frac{1}{2}\nabla_i^2 - \sum_{i,A}\frac{Z_A}{r_{iA}} + \sum_{i>j}\frac{1}{r_{ij}} + \sum_{B>A} \frac{Z_A Z_B}{R_{AB}} \notag\\
-\sum_{A}\frac{1}{2M_A}\nabla_A^2,
\end{align}

where the variables are defined in \cite{PhysRevA.95.042308}. In second quantized form,
\begin{align}
H = \sum_{mn}h_{mn}a^{\dagger}_ma_n + \frac{1}{2}\sum_{mnop} h_{mnop} a^{\dagger}_ma^{\dagger}_na_oa_p,
\end{align}

where $h_{mn}$ and $h_{mnop}$ are evaluated from integrals over the spatial and spin degrees of freedom of electrons. We use the Jordan-Wigner transformation to map the fermionic Hamiltonian into a distinguishable qubit Hamiltonian. The goal of VQE is to find the lowest energy state of a molecule, which is important for understanding its chemical properties and behavior.
The algorithm starts with an initial guess for the wave function of the molecule and then uses a classical optimizer to adjust the parameters of the wave function to minimize the energy. The adjusted wave function is then sent to the quantum computer, which calculates the energy of the wave function. This process is repeated iteratively, with the classical optimizer updating the wave function parameters and the quantum computer calculating the energy until the energy converges to the ground state energy of the molecule. We test the vanilla flavor VQE \cite{https://doi.org/10.48550/arxiv.2103.08505} method; then we explore the Tetris Adaptive VQE \cite{https://doi.org/10.48550/arxiv.2209.10562}, and we finally implement the Adaptive VQE \cite{https://doi.org/10.48550/arxiv.2204.07179, https://doi.org/10.48550/arxiv.2301.10196} method, which gave us the best results.

\begin{figure}[h!]
\centering
\includegraphics[scale=0.5]{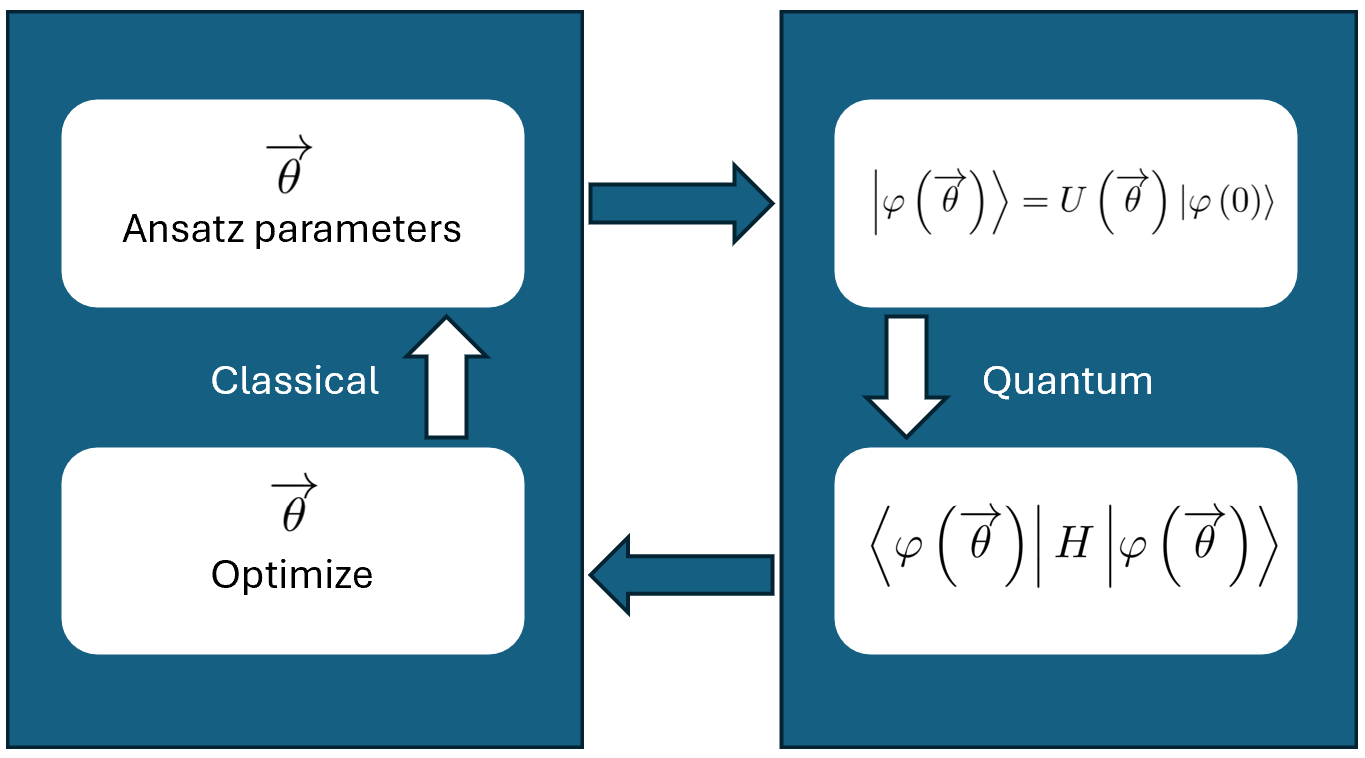}
\caption{VQE is a hybrid algorithm that leverages classical training of parameters on the one hand and evaluation of computationally expensive expectation values on a quantum computer on the other.}
\end{figure}

In addition to the vanilla flavor VQE, we used an active space transformation. This refers to the process of selecting a subset of molecular orbitals and electrons that are crucial for describing the electronic structure of a molecule accurately.
Given a total of $N$ molecular orbitals and $M$ electrons, an active space transformation focuses on a reduced number of orbitals $n$ and electrons $m$. The goal is to simplify the problem while retaining essential correlations:

\[
\text{Active Space Hamiltonian} = P_{\text{active}} H P_{\text{active}}
\]

where $P_{\text{active}}$ is the projection operator onto the active space.



\subsection{ZNE}
In the NISQ era, it's essential to consider error mitigation when trying to run our algorithms on a real quantum hardware. ZNE is an error mitigation technique in quantum computing that can improve the accuracy of quantum algorithms by extrapolating the results of noisy quantum circuits to the limit of zero noise \cite{ZNE}. In a noisy quantum circuit, the presence of unwanted noise can cause errors in the computation, which can reduce the accuracy of the results. ZNE aims to mitigate this problem by extrapolating the noisy results to the limit of zero noise, where the true value of the computation can be obtained without errors. ZNE computes expectation values of observables for different noise factors then uses the measured values to infer the ideal expectation value at the zero-noise limit. Given

\begin{equation}
E_K(\lambda) = E^{*} + \sum_{k=1}^na_k \lambda^k + R_{n+1}(\lambda, \mathcal{L}, T)
\end{equation}

where the variables are defined in \cite{PhysRevLett.119.180509}. ZNE aims to find $\text{lim}_{\lambda \rightarrow 0}E_K(\lambda) = E^{*}$.
One advantage of the ZNE technique is that it can be applied to any quantum algorithm, regardless of its specific structure or purpose. Moreover, ZNE can be combined with other noise mitigation techniques, such as error correction and error suppression, to further improve the accuracy and reliability of quantum computations. However ZNE is not guaranteed to produce unbiased results.

\subsection{TREX}
TREX is a readout error mitigation method that uses twirled noisy expectation values to calculate noise-mitigated expectation values \cite{TREX}. It involves applying a random unitary transformation to the final state of a quantum circuit before measuring it. The idea is to randomly rotate the state space in a way that cancels out certain types of noise, such as systematic errors in the measurement process.

Similar to ZNE, TREX estimates a quantity of the following form in the presence of noise,

\begin{align}
\left \langle Z_b \right \rangle_{\rho} = \text{Tr}\left( Z_b \rho \right) = \sum_{q \in \mathbb{Z}_2^n} (-1)^{\left \langle b, q \right \rangle} \text{Tr}\left(E_q \rho \right)
\end{align}

where the variables are defined in \cite{PhysRevLett.119.180509}. TREX has been shown to be effective in a variety of quantum computing applications, including quantum chemistry simulations, optimization problems, and machine learning tasks. It is a promising technique for overcoming the limitations of noisy, near-term quantum devices and enabling the development of more accurate and reliable quantum algorithms.

\subsection{Circuit Knitting and Entanglement Forging}
Circuit knitting is a technique in quantum computing that involves combining multiple quantum circuits into a larger more complex circuits \cite{Piveteau_Sutter}. Entanglement forging takes a generic circuit that operates on the combined system of the spin-up and spin-down halves and splits it into smaller circuits that only operate on one half at a time \cite{PRXQuantum.3.010309}. In other words, the entanglement forging technique takes a circuit operating on 2$N$ qubits and separates that circuit into two $N$-qubit halves.
The ansatz can be described by a parameterized quantum circuit. Mathematically, a quantum state $\left| \psi(\theta) \right\rangle$ is generated by applying a series of unitary operations $U(\theta)$ on a reference state, typically a Hartree-Fock state:
\[
\left| \psi(\theta) \right\rangle = U(\theta) \left| \psi_{\text{ref}} \right\rangle
\]
Here, $U(\theta)$ represents a unitary operation parameterized by angles $\theta$, and $\left| \psi_{\text{ref}} \right\rangle$ is the initial reference state.
The unitary operation $U(\theta)$ consists of sequences of gate operations designed to encode the complex correlations and entanglement present in the ground state of the system. In the context of the Entanglement Forging Ansatz, $U(\theta)$ is constructed from layers of entangling gates and single-qubit rotations:

\[
U(\theta) = \prod_{k} U_{k}(\theta_{k})
\]

where $U_{k}(\theta_{k})$ represents a unitary gate or a sequence of gates parameterized by $\theta_{k}$.

\section{Results}
\label{sec:results}

\subsection*{Real hardware experiments}
We find the following results from a real quantum device (ibmq-lima) having used error mitigation techniques as described above.
\begin{figure}[h]
\centering
\includegraphics[scale=0.3]{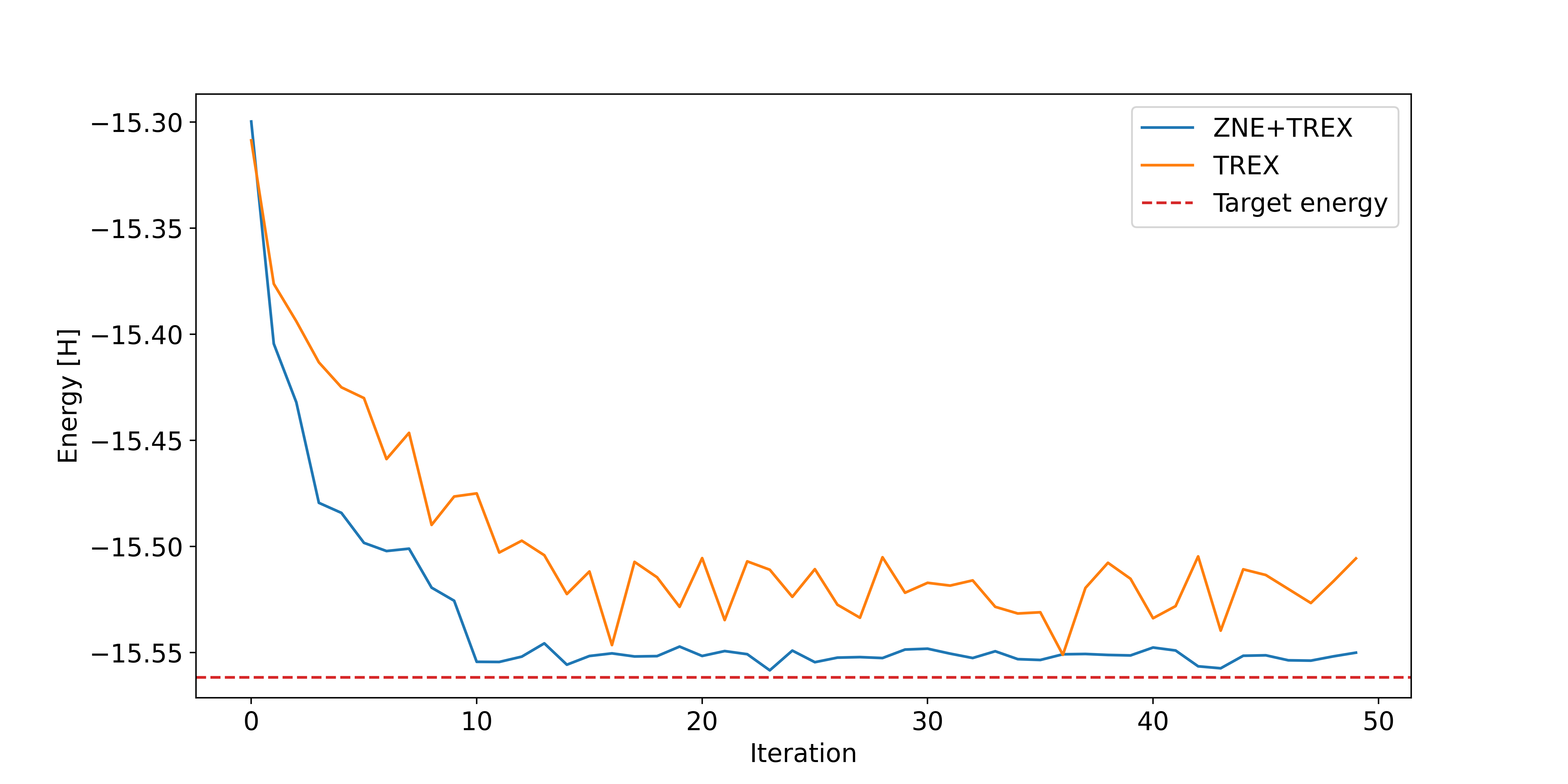}
\caption{VQE results from a real quantum device (ibmq{\_}lima) and from a noisy simulator.}
\end{figure}
\\
We see that using an ensemble method of ZNE+TREX gives us the best convergence, both in terms of accuracy and speed over noisy backend. This provides us with an optimal promise of using variational algorithms and error mitigation techniques to approximate different quantities in Quantum chemistry including the ground state energy of different molecules.

\subsection*{VQE}

First, we do the usual hybrid VQE algorithm using single and double excitations UCCSD strategy \cite{UCCSD}. We use all such excitations to come up with the ground state energy calculation. For this setup, we use 4 electrons in 6 orbitals inside the molecule. The pair of electrons in the 1s orbital (core) of the $Be$ atom is stable. Therefore, we don't include them in our active space. We use the sto-3g basis set and two-qubit reduction method using parity mapping to reduce the system size without affecting the calculation, due to symmetry. We first find the true ground state energy, which we will use throughout the file for reference, as the target. Note that this is calculated using NumpyEigenSolver. We then make a circuit and find the expected value of the hamiltonian using the ADAM optimizer. 
We calculate the time taken for this method, along with the number of CNOT gates that are in the circuit. This number is calculated using the formula 13 $\times$ double{\_}excitations + 2 $\times$ single{\_}excitations. This is because of the number of CNOT gates used in the ansatz for the respective excitation schemes. Finally, we plot the graph of the ground state (GS) Energy calculation over iterations.


\subsection*{Adapt VQE}
In the previous calculation, we used all the single and double excitations. However, one can reduce their number by looking at the gradient w.r.t. parameters for those schemes. The idea is that if the parameters remain constant, then we can train the schemes where parameters need to change in order to reduce the time as well as the number of gates required. The same is demonstrated using adapt VQE, where we `adapt' to non-vanishing gradients and using a similar ansatz with a restricted number of schemes, we get much faster and better results. This is basically due to the classical preprocessing of data. In the VQE algorithm, a parameterized quantum circuit is constructed to prepare a trial state, which is then measured to estimate the expectation value of the Hamiltonian of the system. This expectation value is used to calculate an estimate of the ground state energy of the system, and the parameters of the circuit are iteratively adjusted to minimize the energy. The Adaptive VQE algorithm improves on this by dynamically adjusting the circuit during the optimization process based on the measurement outcomes obtained so far. This allows the algorithm to focus its resources on the most relevant parts of the quantum state space, leading to faster convergence and potentially better accuracy. Adaptive VQE is particularly useful for problems where the Hamiltonian of the quantum system is not known in advance or where it is difficult to simulate classically. It is also a promising approach for implementing quantum machine learning algorithms.

We use a slightly different version of adapt VQE which uses the dynamic gradients twice. We first find the gates corresponding to double excitations with gradient higher than a threshold. After we get the selected double excitations, we fix these double excitations and only look at the gradient of the single excitations in the new circuit. We again cut off the vanishing gradients and look at selected single and selected double excitations for VQE. 
The convergence can be seen after merely 10 iterations, and the time taken is less than a minute for optimization. The number of CNOT gates already reduce by a factor of 10 which is incredible. This provides a great example for NISQ era quantum algorithm that can be implemented successfully.


 \subsection*{Tetris VQE}
The tetris VQE algorithm aims to optimize the circuit depth and entangling structure of the parameterized quantum circuit used in the VQE algorithm, with the goal of reducing the total number of gate operations required to obtain an accurate estimate of the ground state energy of a quantum system. The algorithm achieves this by using a process similar to the game of tetris, where circuit blocks are combined and rearranged to minimize the number of gates required. Tetris VQE can be used for a variety of quantum chemistry applications, such as calculating the ground state energy of small molecules. By reducing the number of gate operations required, tetris VQE has the potential to significantly reduce the computational resources needed for these types of calculations on near-term quantum computers. Overall, tetris VQE is an example of a new wave of algorithmic techniques that are emerging to make better use of the limited resources of current quantum computers and to pave the way for large-scale, fault-tolerant quantum computing in the future.

The following is a circuit for an adapt VQE setup
\hspace*{1.6cm} 
\Qcircuit @C=1em @R=.7em @!R {
  & \lstick{q_0:} 
  & \gate{G_1}  & \multigate{2}{U} & \gate{G_3} & \multigate{1}{V} & \qw \\
  & \lstick{q_1:} 
  &  \qw & \ghost{U} & \qw & \ghost{V} &\gate{G_5}  & \qw \\
  & \lstick{q_2:} 
  & \qw & \ghost{U} &  \multigate{3}{U_2} &\multigate{1}{V_2} & \qw & \qw \\
  & \lstick{q_3:} 
  & \multigate{1}{V} &  \gate{G_2} & \ghost{U_2} &\ghost{V_2} & \multigate{2}{U} & \qw \\
  & \lstick{q_4:} 
  & \ghost{V} & \qw & \ghost{U_2} &\qw& \ghost{U} & \qw \\
  & \lstick{q_5:} 
  & \gate{G_4} & \multigate{1}{W}  &\ghost{U_2} & \qw & \ghost{U} & \qw \\
  & \lstick{q_6:} 
  & \qw & \ghost{W} & \qw & \qw &\qw &\qw \\
}
\\
\\
For this, the tetris adapt VQE circuit could look like:
\\
\hspace*{2.4cm} 
\Qcircuit @C=1em @R=.7em @!R {
  & \lstick{q_0:} 
  & \gate{G_1}  & \multigate{2}{U} & \multigate{1}{V'} & \qw \\
  & \lstick{q_1:} 
  &  \multigate{1}{U'} & \ghost{U}  & \ghost{V'} & \qw \\
  & \lstick{q_2:} 
  & \ghost{U'} & \ghost{U} & \multigate{3}{U_2} & \qw \\
  & \lstick{q_3:} 
  & \multigate{1}{V} &  \multigate{1}{U'} & \ghost{U_2} & \qw \\
  & \lstick{q_4:} 
  & \ghost{V} & \ghost{U'} &  \ghost{U_2} & \qw \\
  & \lstick{q_5:} 
  & \gate{G_4} & \multigate{1}{W}  & \ghost{U_2} & \qw \\
  & \lstick{q_6:} 
  & \gate{G'} & \ghost{W} & \gate{G'} \qw \\
}
\\
\\
The idea is to reduce the depth by possibly adding gates that maximizes the number of the gates (with high gradient) in the given depth. Note that the addition of newer gates need not outperform the adapt VQE version always in a simulation due to the change in circuit. We see something similar in our case, in addition to the fact that we look at the ADAPT circuit using two thresholding instead of one.

\subsection*{Gate complexity}
We calculate the number of two qubit gates used in the three flavors of VQE, namely vanilla, tetris and adapt using the formula mentioned above. 
\begin{table}[h!]
\centering
\begin{tabular}{ |c|c|c| } 
 \hline
 VQE flavor & 2-qubit gates & depth\\
 \hline
 Vanilla VQE & 1020 & 78  \\ 
 TETRIS VQE & 603 & 22 \\ 
 ADAPT VQE & 190 & 15 \\ 
 \hline
\end{tabular}
\caption{Two qubit gate count for different VQE types}
\label{table: 2-qubit count}
\end{table}
\\
We see that adapt VQE has the lowest CNOT count. Furthermore, we see that adapt VQE converges better than the other flavors of VQE with a higher precision. 

\begin{figure}[h]
\centering
\includegraphics[scale=0.3]{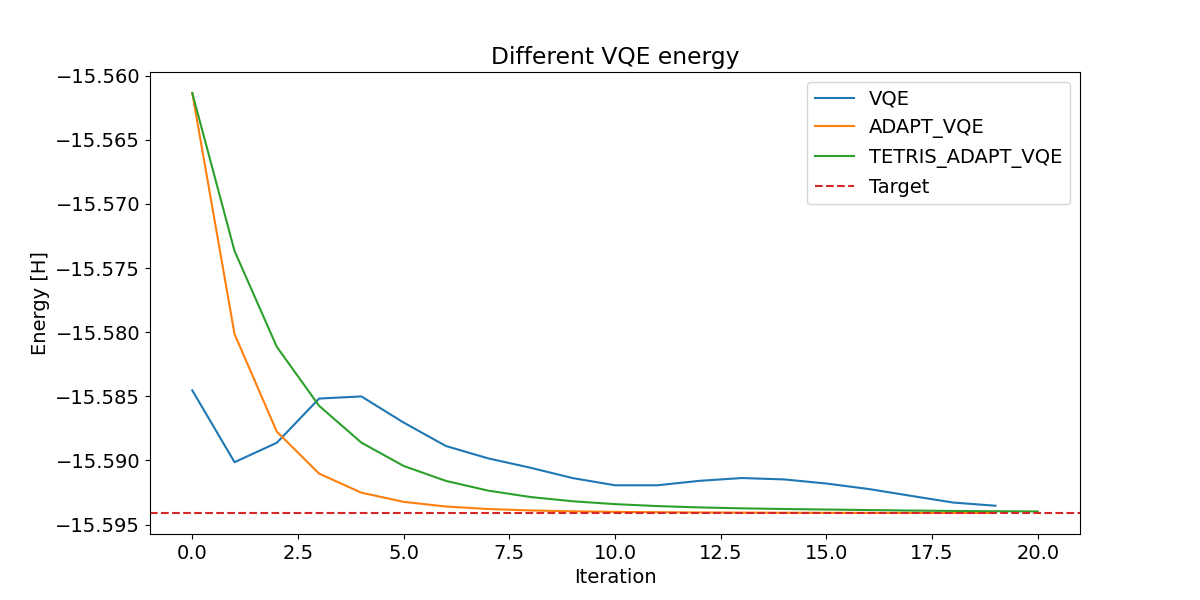}
\caption{Using different VQE algorithms. Our implementation of the adapt VQE method converges fastest to the true ground state of $BeH_2$ compared to our implementation of VQE and tetris-adaptive VQE.}
\end{figure}

We now present results from using entanglement forging on a subsystem of $BeH_2$.

\begin{figure}[h]
\centering
\includegraphics[scale=0.4]{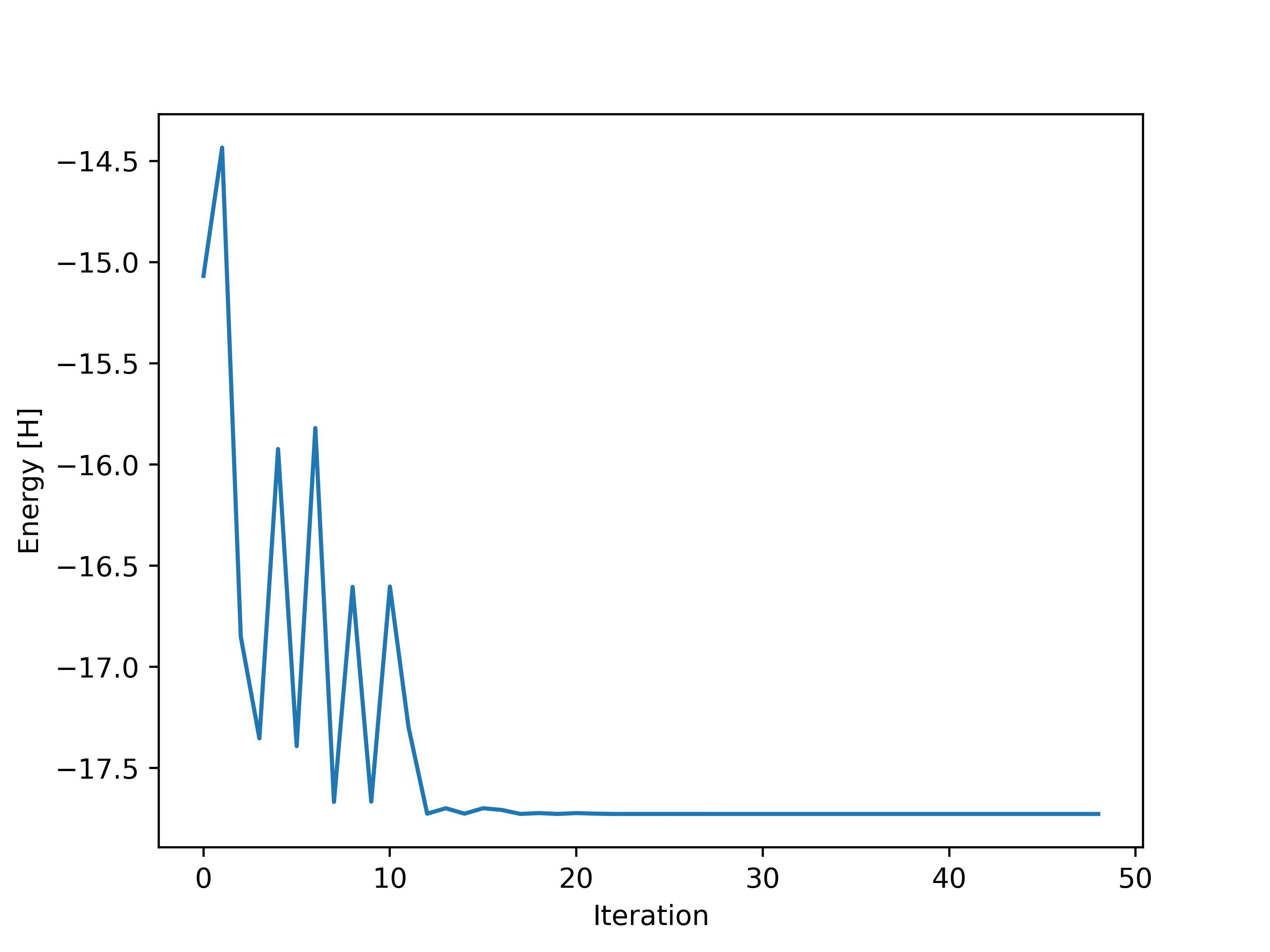}
\caption{Using entanglement forging. The graph shows that using a subsystem of two active orbitals loses chemical accuracy.}
\end{figure}



\section{Discussion}
\label{sec:discussion}
We finally discuss results from the different the VQE methods and various error mitigation techniques we used. The GS energy values for the $BeH_2$ using different molecules were found to converge to the expected value of -15.5943 Hartrees. Here, the adapt VQE method gives the best results, but we need to analyze the case on a noisy simulator as well, and this is what we provide. Here, we also provide what GS values we get employing ZNE. The results show that the qiskit runtime error mitigation works quite well. Additionally, we have error-mitigated results from the real backend ibmq{\_}lima using TREX and ZNE. While TREX results show monotonic convergence to the ground state, we show ZNE results. In conclusion, we provide different methodologies for ground state computation on noiseless and noisy backends (including the real backend). We see adapt VQE works best for the noiseless backend. The double thresholding method can be extended to the tetris-adapt VQE to see if we can reduce the depth even further. For real quantum devices, we used error mitigation techniques of TREX and ZNE to realize that we get very close to the actual answer, which is incredible even after using a subsystem. For the entanglement forging results, the number of orbitals has been oversimplified to two losing chemical accuracy while increasing computational speed. The same process can be carried out for more accuracy with more orbitals and picking appropriate bitstrings given the molecular electronic distribution. The number of 1s in each bitstring must equal the number of 
spin up or spin down electrons and the number of bits in each bitstring should equal the number of spatial orbitals.

\section{Conclusion}
\label{sec:conclusion}
The research uses quantum computing techniques to investigate and compare various VQE methods and error mitigation strategies. Our primary focus was on the computation of GS energy values for the $BeH_2$ molecule. The results demonstrated a convergence to the anticipated value of -15.5943 Hartrees, with the adapt VQE method emerging as the most effective in a noiseless environment.
However, the complexity of real-world quantum computing necessitates the extension of this analysis to noisy simulators, an aspect we have addressed in our work. We have incorporated the use of ZNE and presented the GS values obtained through this technique. The findings indicate a commendable performance of the Qiskit runtime error mitigation.
Furthermore, we ventured into the realm of real quantum devices, employing the backend ibmq{\_}lima. We have utilized error mitigation techniques such as TREX and ZNE. The TREX results exhibited a monotonic convergence to the ground state, and we have also presented the outcomes using ZNE.

Overall, this research provides a robust overview of different methodologies for ground state computation, encompassing both noiseless and noisy backends, including real backends. Our findings suggest that our version of adapt VQE offers optimal performance for noiseless backends. For real quantum devices, we have employed error mitigation techniques of TREX and ZNE, which have yielded results remarkably close to the actual answer, even when utilizing a subsystem. Entanglement forging results over a subsystem sacrifice chemical accuracy but gain speed.
This research significantly contributes to the field of quantum computing and its applications, specifically in the realm of quantum chemistry. It paves the way for future explorations and advancements, with the intention to conduct more robust experiments with efficient code implementation. We also aim to apply the techniques used here to investigate problem-inspired circuit ansatzes and compare them with hardware-efficient ansatzes in subsequent work. Further future prospects include incorporating quantum error correction \cite{arXiv:2208.08571} and exploring hybrid quantum algorithms in the context of time-dependent expectation values \cite{PhysRevLett.121.170501, PhysRevB.108.134301}.

\section*{Data Availability}
The code and data for this work is available at
\url{https://github.com/jsaroni/QHack2023_Feynmanprodigies}.

\section*{Acknowledgements}
We acknowledge the IBM Quantum hardware provided to us during QHack and the PennyLane platform. We would like to thank Xanadu for organizing QHack and giving us the opportunity to work on this project. J.S. thanks IBM Research at Yorktown Heights.

\bibliographystyle{IEEEtran}
\bibliography{refs}

\begin{thebibliography}{10}
\providecommand{\url}[1]{#1}
\csname url@samestyle\endcsname
\providecommand{\newblock}{\relax}
\providecommand{\bibinfo}[2]{#2}
\providecommand{\BIBentrySTDinterwordspacing}{\spaceskip=0pt\relax}
\providecommand{\BIBentryALTinterwordstretchfactor}{4}
\providecommand{\BIBentryALTinterwordspacing}{\spaceskip=\fontdimen2\font plus
\BIBentryALTinterwordstretchfactor\fontdimen3\font minus
  \fontdimen4\font\relax}
\providecommand{\BIBforeignlanguage}[2]{{%
\expandafter\ifx\csname l@#1\endcsname\relax
\typeout{** WARNING: IEEEtran.bst: No hyphenation pattern has been}%
\typeout{** loaded for the language `#1'. Using the pattern for}%
\typeout{** the default language instead.}%
\else
\language=\csname l@#1\endcsname
\fi
#2}}
\providecommand{\BIBdecl}{\relax}
\BIBdecl

\bibitem{10.1038/nature23879}
\BIBentryALTinterwordspacing
A.~Kandala, A.~Mezzacapo, K.~Temme, M.~Takita, M.~Brink, J.~M. Chow, and J.~M.
  Gambetta, ``Hardware-efficient variational quantum eigensolver for small
  molecules and quantum magnets,'' \emph{Nature}, vol. 549, p. 134301, Sep
  2017. [Online]. Available: \url{https://www.nature.com/articles/nature23879}
\BIBentrySTDinterwordspacing

\bibitem{PhysRevA.95.042308}
\BIBentryALTinterwordspacing
J.~R. McClean, M.~E. Kimchi-Schwartz, J.~Carter, and W.~A. de~Jong, ``Hybrid
  quantum-classical hierarchy for mitigation of decoherence and determination
  of excited states,'' \emph{Phys. Rev. A}, vol.~95, p. 042308, Apr 2017.
  [Online]. Available:
  \url{https://link.aps.org/doi/10.1103/PhysRevA.95.042308}
\BIBentrySTDinterwordspacing

\bibitem{https://doi.org/10.48550/arxiv.2103.08505}
\BIBentryALTinterwordspacing
D.~A. Fedorov, B.~Peng, N.~Govind, and Y.~Alexeev, ``Vqe method: A short survey
  and recent developments,'' 2021. [Online]. Available:
  \url{https://arxiv.org/abs/2103.08505}
\BIBentrySTDinterwordspacing

\bibitem{https://doi.org/10.48550/arxiv.2209.10562}
\BIBentryALTinterwordspacing
P.~G. Anastasiou, Y.~Chen, N.~J. Mayhall, E.~Barnes, and S.~E. Economou,
  ``Tetris-adapt-vqe: An adaptive algorithm that yields shallower, denser
  circuit ansätze,'' 2022. [Online]. Available:
  \url{https://arxiv.org/abs/2209.10562}
\BIBentrySTDinterwordspacing

\bibitem{https://doi.org/10.48550/arxiv.2204.07179}
\BIBentryALTinterwordspacing
H.~R. Grimsley, G.~S. Barron, E.~Barnes, S.~E. Economou, and N.~J. Mayhall,
  ``Adapt-vqe is insensitive to rough parameter landscapes and barren
  plateaus,'' 2022. [Online]. Available: \url{https://arxiv.org/abs/2204.07179}
\BIBentrySTDinterwordspacing

\bibitem{https://doi.org/10.48550/arxiv.2301.10196}
\BIBentryALTinterwordspacing
C.~Feniou, M.~Hassan, D.~Traoré, E.~Giner, Y.~Maday, and J.-P. Piquemal,
  ``Overlap-adapt-vqe: Practical quantum chemistry on quantum computers via
  overlap-guided compact ansätze,'' 2023. [Online]. Available:
  \url{https://arxiv.org/abs/2301.10196}
\BIBentrySTDinterwordspacing

\bibitem{ZNE}
\BIBentryALTinterwordspacing
T.~Giurgica-Tiron, Y.~Hindy, R.~LaRose, A.~Mari, and W.~J. Zeng, ``Digital zero
  noise extrapolation for quantum error mitigation,'' in \emph{2020 IEEE
  International Conference on Quantum Computing and Engineering (QCE)}.\hskip
  1em plus 0.5em minus 0.4em\relax Los Alamitos, CA, USA: IEEE Computer
  Society, oct 2020, pp. 306--316. [Online]. Available:
  \url{https://doi.ieeecomputersociety.org/10.1109/QCE49297.2020.00045}
\BIBentrySTDinterwordspacing

\bibitem{PhysRevLett.119.180509}
\BIBentryALTinterwordspacing
K.~Temme, S.~Bravyi, and J.~M. Gambetta, ``Error mitigation for short-depth
  quantum circuits,'' \emph{Phys. Rev. Lett.}, vol. 119, p. 180509, Nov 2017.
  [Online]. Available:
  \url{https://link.aps.org/doi/10.1103/PhysRevLett.119.180509}
\BIBentrySTDinterwordspacing

\bibitem{TREX}
E.~Berg, Z.~Minev, and K.~Temme, ``Model-free readout-error mitigation for
  quantum expectation values,'' \emph{Physical Review A}, vol. 105, 03 2022.

\bibitem{Piveteau_Sutter}
C.~Piveteau and D.~Sutter, ``Circuit knitting with classical communication,''
  \emph{IEEE Transactions on Information Theory}, vol.~PP, pp. 1--1, 01 2023.

\bibitem{PRXQuantum.3.010309}
\BIBentryALTinterwordspacing
A.~Eddins, M.~Motta, T.~P. Gujarati, S.~Bravyi, A.~Mezzacapo, C.~Hadfield, and
  S.~Sheldon, ``Doubling the size of quantum simulators by entanglement
  forging,'' \emph{PRX Quantum}, vol.~3, p. 010309, Jan 2022. [Online].
  Available: \url{https://link.aps.org/doi/10.1103/PRXQuantum.3.010309}
\BIBentrySTDinterwordspacing

\bibitem{UCCSD}
\BIBentryALTinterwordspacing
R.~Xia and S.~Kais, ``Qubit coupled cluster singles and doubles variational
  quantum eigensolver ansatz for electronic structure calculations,''
  \emph{Quantum Science and Technology}, vol.~6, no.~1, p. 015001, oct 2020.
  [Online]. Available: \url{https://dx.doi.org/10.1088/2058-9565/abbc74}
\BIBentrySTDinterwordspacing

\bibitem{arXiv:2208.08571}
T.~Pandey, ``{Topological phases of matter, Quantum Error Correction and
  Topological twist},'' Aug 2022.

\bibitem{PhysRevLett.121.170501}
\BIBentryALTinterwordspacing
H.~Lamm and S.~Lawrence, ``Simulation of nonequilibrium dynamics on a quantum
  computer,'' \emph{Phys. Rev. Lett.}, vol. 121, p. 170501, Oct 2018. [Online].
  Available: \url{https://link.aps.org/doi/10.1103/PhysRevLett.121.170501}
\BIBentrySTDinterwordspacing

\bibitem{PhysRevB.108.134301}
\BIBentryALTinterwordspacing
J.~Saroni, H.~Lamm, P.~P. Orth, and T.~Iadecola, ``Reconstructing thermal
  quantum quench dynamics from pure states,'' \emph{Phys. Rev. B}, vol. 108, p.
  134301, Oct 2023. [Online]. Available:
  \url{https://link.aps.org/doi/10.1103/PhysRevB.108.134301}
\BIBentrySTDinterwordspacing

\end{thebibliography}

\end{document}